\documentclass[12pt]{iopart}

\begin{document}

\title{Paramagnetism in color superconductivity and compact stars}
\author{Efrain J. Ferrer and Vivian de la Incera}
\address{Department of Physics, Western Illinois University, Macomb,
IL 61455, USA}

\begin{abstract} It is quite plausible that color superconductivity occurs in the
inner regions of neutron stars. At the same time, it is known that
strong magnetic fields exist in the interior of these compact
objects. In this paper we discuss some important effects that can
occur in the color superconducting core of compact stars due to the
presence of the stars' magnetic field. In particular, we consider
the modification of the gluon dynamics for a color superconductor
with three massless quark flavors in the presence of an external
magnetic field. We show that the long-range component of the
external magnetic field that penetrates the color-flavor locked
phase produces an instability for field values larger than the
charged gluons' Meissner mass. As a consequence, the ground state is
restructured forming a vortex state characterized by the
condensation of charged gluons and the creation of magnetic flux
tubes. In the vortex state the magnetic field outside the flux tubes
is equal to the applied one, while inside the tubes its strength
increases by an amount that depends on the amplitude of the gluon
condensate. This paramagnetic behavior of the color superconductor
can be relevant for the physics of compact stars.
\end{abstract}
\pacs{12.38.Aw, 12.38.-t, 24.85.+p}
\submitto{\JPA}
\maketitle

\section{Introduction}

In the realm of high density and low temperature QCD baryons get so
squeezed that they start to overlap, thereby erasing any vestige of
structure. Since in that situation the quarks get very close to each
other, their interactions become weak due to asymptotic freedom. At
densities of the order of $10$ times the nuclear density $(\sim
2-4\times 10^{15} g/cm^3)$ the weakly interacting quarks can exist
out of confinement. In nature the combination of such densities and
relatively low temperatures exist in the core of neutron stars,
which are the remnant of supernova explosions. It has been predicted
on purely theoretical grounds that if the remnant of a supernova
explosion has sufficiently high density, it could lead to the
formation of a quark star \cite{Witten}. This compact object would
be something in between a neutron star and a black hole.

At very low temperatures a finite density of fermions will fill out
all the lowest available energy states up to the Fermi energy.
Fermions in the Fermi surface have the same energy, but different
momentum. If there were no attractive interaction between the
fermions sitting on the Fermi surface, this realization will be the
system final state. However, an arbitrarily weak attractive
interaction among those fermions will render the existing ground
state unstable favoring the formation of fermion-fermion pairs. This
restructuring of the ground state is the basis of the phenomena of
superconductivity and superfluidity.

In QCD the fundamental interaction between two quarks is attractive.
Hence, at very large densities the arbitrarily weak interaction
between the asymptotically free quarks on the Fermi surface will do
the trick of restructuring the ground state through the formation of
Cooper pairs of quarks with opposite spin and momentum \cite{CS}.
Because the quarks carry "color" charge, the quark-quark pairs will
carry nonzero color charge too, thus the name of color
superconductivity.

On the other hand, it is well-known that strong magnetic fields, as
large as $B \sim 10^{12} - 10^{14}$ G,  exist in the surface of
neutron stars \cite{Grasso}, while in magnetars they are in the
range $B \sim 10^{14} - 10^{15}$ G, and perhaps as high as $10^{16}$
G \cite{magnetars}. It is presumed from the virial theorem
\cite{virial} that the interior field in neutron stars could be as
high as $10^{18}-10^{19}$ G. If quark stars are self-bound rather
than gravitational-bound objects, the previous upper limit that has
been obtained by comparing the magnetic and gravitational energies
could go even higher. Thus, investigating the effect of strong
magnetic fields in color superconductivity is of interest for the
study of compact stars in astrophysics.

To consider the magnetic field interaction with the particles
immersed in the color superconductor (CS) we should have in mind
that there the quark-quark pairs carry both color and electric
charges. Hence a CS is also an electric superconductor. One might
think that because of this, a magnetic field cannot penetrate the
color superconductor. But in this complex medium something
qualitatively new takes place; the electromagnetic field mixes up
with one of the gluons to form a new "electromagnetic" field (called
in the literature a "rotated" electromagnetic field, where the
"rotation" takes place here in an inner space)
\cite{alf-raj-wil-99/537}. This "rotated" electromagnetic field
$\widetilde{A}$ remains long-range within the superconductor,
because the quark pairs are all neutral with respect to the
corresponding "rotated" electromagnetic charge $\widetilde{Q}$.
Therefore, there is no Meissner effect for the corresponding rotated
magnetic field $\widetilde{H}$.

In recent works \cite{MCFL, Vortex} we showed that the properties of
the CS can be substantially transformed by the penetrating
$\widetilde{H}$ field. First, the pairing of (rotated) electrically
charged quarks is reinforced by the field \cite{MCFL}. Pairs of this
kind have bounding energies which depend on the magnetic-field
strength and are bigger than the ones existing at zero field.
Second, the symmetry of the superconducting phase is changed,
because now the magnetic field differentiates the condensates which
get contributions from pairs formed by $\widetilde{Q}$-charged
quarks from those that only get contributions from pairs formed by
$\widetilde{Q}$-neutral quarks \cite{MCFL}. Due to the symmetry
change, the low-energy physics of the superconductor is also
changed. This last effect can have practical implications for
astrophysics since all the transport properties of the star are
basically managed by the low-energy physics of the phase. In
particular, the cooling of the star is determined by the particles
with the lowest energy; so a star with a core of quark matter and a
sufficiently large magnetic field can have a distinctive cooling
process. This is a point that deserves to be investigated in more
detail. Finally, the magnetic field can also influence the gluon
dynamics \cite{Vortex}. At field strengths comparable to the charged
gluon Meissner mass an inhomogeneous condensate of
$\widetilde{Q}$-charged gluons is formed \cite{Vortex}. The gluon
condensate anti-screens the magnetic field due to the anomalous
magnetic moment of these spin-1 particles. Because of the
anti-screening, this condensate does not give a mass to the
$\widetilde{Q}$ photon, but instead amplifies the applied rotated
magnetic field. This means that in the CS a sort of anti-Meissner
effect takes place. This last effect can be also of interest for
astrophysics since once the core of a compact star becomes color
superconducting, its internal magnetic field can be boosted to
values higher than those found in neutron stars with cores of
nuclear matter. This effect could open a new window to differentiate
a neutron star made up entirely of nuclear matter from one with a
quark matter core. In this paper we will discuss the mechanism that
generates this kind of paramagnetism in color superconductivity.

\section{Gluon instability at $\widetilde{H}\geq\widetilde{H}_{C}$}

There is a similarity between the electroweak symmetry-broken phase
and the color superconducting phase of QCD. In the first model, the
Higgs condensate although blocks the penetration of the
hypermagnetic field, allows a combination of the hyperfield and one
of the weak isospin fields to penetrate the symmetry-broken medium.
As known, the corresponding penetrating field is the electromagnetic
field $A$, which is the only remaining long-range field in that
phase. The $W^{\pm}$ bosons, although neutral with respect to the
hypercharge, acquire electromagnetic charges in the new phase. In
the CS, the role of the electromagnetic field is played by the
linear combination
$\widetilde{A}_{\mu}=\cos{\theta}\,A_{\mu}+\sin{\theta}\,G^{8}_{\mu}$
of the photon $A_{\mu}$ and the gluon $G^{8}_{\mu}$ fields. Even
though gluons are neutral with respect to the conventional
electromagnetism, in the color superconducting phase they acquire
$\widetilde{Q}$ charges:
\begin{equation} \label{table}
\begin{tabular}{|c|c|c|c|c|c|c|c|c|}
  \hline
  $G_{\mu}^{1}$ & $G_{\mu}^{2}$ & $G_{\mu}^{3}$ & $G_{\mu}^{+}$ & $G_{\mu}^{-}$ & $I_{\mu}^{+}$ & $I_{\mu}^{-}$ & $\widetilde{G}_{\mu}^{8}$ \\
  \hline
  0 & 0 & 0 & 1 & -1 & 1 & -1 & 0 \\
  \hline
\end{tabular} \ ,
\end{equation}
given in units of $\widetilde{e} = e \cos{\theta}$. The
$\widetilde{Q}$-charged fields in (\ref{table}) correspond to the
combinations $G_{\mu}^{\pm}\equiv\frac{1}{\sqrt{2}}[G_{\mu}^{4}\mp
iG_{\mu}^{5}]$ and
$I_{\mu}^{\pm}\equiv\frac{1}{\sqrt{2}}[G_{\mu}^{6}\mp
iG_{\mu}^{7}]$.

Taking into account the Schwinger energy spectrum of a charged
particle of spin $s$, charge $e$, gyromagnetic ratio $g$, and mass
$m$ in a magnetic field $H$,

\begin{equation}
\label{Sch-Eq} E^{2}_{n}=(2n+1)eH-ge\textbf{H}\cdot
\textbf{s}+m^{2},
\end{equation}
we see that for spin-1 particles (i.e. $g=2$ and spin projection -1,
0, +1) $E^{2}<0$ for strong enough magnetic fields ($H >
H_{cr}=m^{2}/e$). Therefore, when the field surpasses the critical
value $H_{cr}$, one of the modes of the charged gauge field becomes
tachyonic (this is the well known "zero-mode problem" found in the
presence of a magnetic field for Yang-Mills fields \cite{zero-mode},
for the $W^{\pm}_{\mu}$ bosons in the electroweak theory
\cite{Skalozub, Olesen}, and even for higher-spin fields in the
context of string theory \cite{porrati}).

Similarly to other spin-1 theories with magnetic instabilities
\cite{zero-mode}-\cite{Olesen}, the charged gluons in the magnetized
CS suffer of instabilities for $\widetilde{H} >
\widetilde{H}_{C}=m_{M}^{2}/\widetilde{e}$, where $m_{M}$ is the
charged gluon Meissner mass. To remove the magnetically induced
instabilities, a vortex ground state is formed \cite{Vortex}. This
vortex state is characterized by the condensation of charged gluons
and the creation of "rotated" magnetic flux tubes.

\section{Gluon vortex condensate and paramagnetism}

Since at densities high enough to neglect the strange quark mass,
the ground state of three-flavor quark matter corresponds to the
color-flavor locked ($CFL$) phase \cite{alf-raj-wil-99/537}, we will
focus our analysis into this phase, although the conclusions can be
easily extrapolated to other phases, as the $2SC$ phase, for
example.

Above the critical field ($\widetilde{H}_{C}=
m_{M}^2/\widetilde{e}$) oriented along the $Z$-axis, the mass mode
that becomes tachyonic, corresponds to a charged field eigenvector
of amplitude $G$ in the $(1,i)$ spatial $(x,y)$-direction for
$G^{-}$ ($G^{\ast}$ in the $(1,-i)$ direction for $G^{+}$). Without
loss of generality, we are only considering in this analysis the set
of charged fields $G_{\mu}^{\pm}$. To remove the tachyonic mode, the
ground state is restructured through the formation of a gauge field
condensate $G$, as well as an induced magnetic field
$\widetilde{\textbf{B}}=\nabla\times\widetilde{\textbf{A}}$ that is
originated due to the backreaction of the G condensate on the
rotated electromagnetic field.

The condensate solutions can be found by minimizing with respect to
$G$ and $\widetilde{B}$ the Gibbs free energy density
$\mathcal{G}_{c}=\mathcal{F}-\widetilde{H}\widetilde{B}$,
($\mathcal{F}$ is the free energy density)

\begin{eqnarray}
\label{Gibbs} \mathcal{G}_{c} =\mathcal{F}_{n0}
-2G^{\dag}\widetilde{\Pi}^{2}
G-2(2\widetilde{e}\widetilde{B}-m_{M}^{2})|G|^{2}+2g^{2}|G|^{4}\nonumber
 \\
+ \frac{1}{2}\widetilde{B}^{2}-\widetilde{H}\widetilde{B}\qquad
\qquad \qquad \qquad \qquad \qquad \qquad
\end{eqnarray}
In (\ref{Gibbs}) $\mathcal{F}_{n0}$ is the system free energy
density in the normal-CFL phase ($G=0$) at zero applied field. Using
(\ref{Gibbs}) the minimum equations at $\widetilde{H}\sim
\widetilde{H}_{C}$ for the condensate $G$ and induced field
$\widetilde{B}$ respectively are
\begin{equation}
\label{G-Eq} -\widetilde{\Pi}^{2}
G-(2\widetilde{e}\widetilde{B}-m_{M}^{2})G=0,
\end{equation}
\begin{equation}
\label{B-Eq} 2\widetilde{e} |G|^{2}-\widetilde{B}+\widetilde{H}=0
\end{equation}
Identifying $G$ with the complex order parameter, Eqs.
(\ref{G-Eq})-(\ref{B-Eq}) become analogous to the Ginzburg-Landau
equations for a conventional superconductor except for the
$\widetilde{B}$ contribution in the second term in (\ref{G-Eq}) and
the sign of the first term in (\ref{B-Eq}). The origin of both terms
can be traced back to the anomalous magnetic moment term
$4\widetilde{e}\widetilde{B}|G|^{2}$ in the Gibbs free energy
density (\ref{Gibbs}). Notice that because of the different sign in
the first term of (\ref{B-Eq}), contrary to what occurs in
conventional superconductivity, the resultant field $\widetilde{B}$
is stronger than the applied field $\widetilde{H}$. Thus, when a
gluon condensate develops, the magnetic field will be antiscreened
and the CS will behave as a paramagnet. The antiscreening of a
magnetic field has been also found in the context of the electroweak
theory for magnetic fields $H \geq M_{W}^{2}/e\sim 10^{24} G$
\cite{Olesen}. Just as in the electroweak case, the antiscreening in
the CS is a direct consequence of the asymptotic freedom of the
underlying theory \cite{Olesen, Hughes}.

We should highlight that the gluon condensate discussed in this work
is not the only charged spin-one condensate generated in a theory
with a large fermion density. As known \cite{Linde}, a spin-one
condensate of $W^{\pm}$-bosons can be originated at sufficiently
high fermion density in the context of the electroweak theory at
zero magnetic field. However, the physical implications of the gluon
condensate induced by the magnetic field in the CS are fundamentally
different from those associated to the homogeneous $W^{\pm}$-boson
condensate of the dense electroweak theory \cite{Linde}. The gluon
vortices in the magnetized CS boost the applied field, leaving the
$\widetilde{Q}$ photon massless and thereby preserving the
$\widetilde{U}_{em}(1)$ symmetry. On the other hand, the
$W^{\pm}$-boson condensate breaks the $U_{em}(1)$ symmetry turning
the electroweak system in an electromagnetic superconductor
\cite{Shabad}.

The explicit solution of (\ref{G-Eq}) with vanishing conditions at
$x\rightarrow\pm\infty$, can be found following Abrikosov's approach
\cite{Abrikosov} to type II metal superconductivity for the limit
situation when the applied field is near the critical value
$H_{c2}$. In our case we find

\begin{equation}
\label{Sol-3} G_{k}=\exp {[-iky]}
\exp{[-\frac{(x-x_{k})^2}{2\xi^{2}}]}
\end{equation}
where $k\equiv k_{y}$. From the experience with conventional type II
superconductivity  \cite{Tinkham} it is known that to minimize the
energy the inhomogeneous condensate solutions secure a periodic
lattice structure. Then, putting on periodicity in the $y$-direction
with period $\Delta y= b$ restricts the values of $k$ to a discrete
set $k=2\pi n/b$,   $n=1,2,..$. This condition implies that we have
an infinite set of discrete solutions that superpose to form the
general solution $G(x,y)=\sum C_{n}G_{n}$. This superposition of all
the Gaussian solutions centered at different $x_{n}$ constitutes the
vortex state that removes the instability in the whole space. On the
other hand, the discrete values of $k$ imply periodicity in $x$,
since the Gaussian solutions $G_{n}$ are located at
$x_{n}=\frac{k_{n}\widetilde{\Phi}_{0}}{2\pi
\widetilde{H}_{C}}$=$\frac{ n\widetilde{\Phi}_{0}}{b
\widetilde{H}_{C}}$, with
$\widetilde{\Phi}_{0}\equiv2\pi/\widetilde{e}$. Assuming then that
all $G_{n}$ enter with equal weight, the periodicity length in the
$x$-direction is $\Delta x=\frac{\widetilde{\Phi}_{0}}{b
\widetilde{H}_{C}}$. Therefore, the magnetic flux through each
periodicity cell in the vortex lattice is quantized $\label{Flux}
\widetilde{H}_{C}\Delta x \Delta y=\widetilde{\Phi}_{0}$, with
$\widetilde{\Phi}_{0}$ being the flux quantum per unit vortex cell.
In this semi-qualitative analysis we considered Abrikosov's ansatz
of a rectangular lattice (i.e. all the coefficients $C_{n}$ being
equal). For the rectangular lattice, the area of the unit cell is
$A=\Delta x \Delta y=\widetilde{\Phi}_{0} /\widetilde{H}_{C}$, so
decreasing with $\widetilde{H}$.

\section{Conclusions and Final Remarks}

In conclusion, at low $\widetilde{H}$ field the CS behaves as an
insulator that can be penetrated by the $\widetilde{H}$ field. When
the $\widetilde{H}$ field reaches the critical value
$\widetilde{H}_{C}= m_{M}^2/\widetilde{e}$, the condensation of
charged gluons is triggered inducing the formation of a lattice of
magnetic flux tubes and breaking the translational and remaining
rotational symmetries. Contrary to the situation in conventional
type-II superconductors, where the applied field only penetrates
through the flux tubes and with a smaller strength, the vortex state
in the CS has the peculiarity that outside the flux tube the applied
field $\widetilde{H}$ totally penetrates the sample, while inside
the tubes the magnetic field becomes larger than $\widetilde{H}$.
This antiscreening behavior is similar to that of the electroweak
system at high magnetic field \cite{Olesen}. Notice that as the
$\widetilde{Q}$ photons remain massless in the presence of the
condensate $G$, the $\widetilde{U}(1)_{em}$ symmetry remains
unbroken.

A rough estimate of the critical field that produces the magnetic
instability at the scale of baryon densities typical of neutron-star
cores ($\mu \simeq 200-400 MeV$, $\alpha_{s}(\mu) \simeq 1/3$) gives
$\widetilde{H}_{C}\simeq 9.5\times 10^{16}G-3.8\times 10^{17}G$.
Although these are significantly high magnetic fields, they cannot
be ruled out as acceptable values for the neutron star core.

At present, there is a lot of activity among the physics community
trying to find ways to differentiate a neutron star made up entirely
of nuclear matter from that with a quark color superconducting core.
Some guiding ideas in this direction have been to link the phase of
the star's core to measurable properties of the star as its
radio-mass ratio, its cooling process, and its rotational and
vibrational properties. In this regard, the result that we are
reporting on the increase of the star's magnetic field by the
realization of color superconductivity in its core can also serve to
that goal, and deserves further investigations.

\ack

This work has been supported in part by DOE grant DE-FG02-07ER41458.

\end{document}